\documentclass[twocolumn]{revtex4}

\usepackage{graphicx}

\begin{document}


\title{Loschmidt echo in a system of interacting electrons}
\author{G. Manfredi}
\email{Giovanni.Manfredi@ipcms.u-strasbg.fr}
\author{P.-A. Hervieux}
\affiliation{Institut de Physique et Chimie des Mat{\'e}riaux de
Strasbourg \\ UMR 7504 ULP-CNRS, 23 rue du Loess, BP 43, F-67034
Strasbourg Cedex 2, France}

\date{\today}
\begin{abstract}
We study the Loschmidt echo for a system of electrons interacting
through mean-field Coulomb forces. The electron gas is modeled by
a self-consistent set of hydrodynamic equations. It is observed
that the quantum fidelity drops abruptly after a time that is
proportional to the logarithm of the perturbation amplitude. The
fidelity drop is related to the breakdown of the symmetry
properties of the wave function.
\end{abstract}

\maketitle

{\it Introduction}.---In a famous controversy with L. Boltzmann,
J. Loschmidt pointed out that, if one reverses the velocities of
all particles in a physical system, the latter would evolve back
to its initial state, thus violating the second law of
thermodynamics. The main objection to this argument is that
velocity reversal is a very unstable operation and tiny errors in
the reversal quickly restore normal entropy increase.

More recently, the original idea of Loschmidt was revived in the
context of quantum information theory. Indeed, any attempt at
coding information using quantum bits is prone to failure if a
small coupling to an uncontrollable environment destroys the
unitary evolution of the wave function (decoherence) \cite{zurek}.
In order to estimate the robustness of the system against
perturbations from the environment, the following procedure has
been suggested. The system is allowed to evolve under the action
of an unperturbed Hamiltonian until time $T$; then it is evolved
backwards in time until $2T$ with the original Hamiltonian plus a
small perturbation (the `environment'). The square of the scalar
product of the initial and final states defines the quantum
fidelity of the system (Loschmidt echo) and has been the object of
intense study in recent years. Jalabert and Pastawski
\cite{jalabert} have proven that, for perturbations that are
classically weak but quantum-mechanically strong, the fidelity
decay rate only depends on the classical Lyapunov exponent of the
unperturbed system. This universal behavior was later corroborated
by numerical simulations \cite{jacquod,cucchietti} and experiments
\cite{echo-exp}. For weaker perturbations, the decay rate is still
exponential, but perturbation-dependent (Fermi golden rule
regime). For still weaker perturbations, the decay is Gaussian
\cite{jacquod}.

An equivalent approach to the Loschmidt echo was proposed earlier
by Peres \cite{peres}. In order to study the separation of
classical trajectories, it is customary to compare the evolution
of two slightly different initial conditions. Peres noted that one
could just as well compare the {\em same} initial condition
evolving in two slightly different Hamiltonians, an unperturbed
one $H_0$ and a perturbed one $H=H_0+\delta H$. The fidelity at
time $t$ is then defined as the square of the scalar product of
the wave functions evolving with $H_0$ and $H$ respectively: $F(t)
= ~\vline \langle \psi_{H_0}(t) \vline ~\psi_H(t) \rangle
\vline^{~2}$. The latter approach is the one adopted throughout
the present paper.

Virtually all theoretical investigations of the Loschmidt echo
consider one-particle systems evolving in a given (usually
chaotic) Hamiltonian. The aim of the present work is to explore
the more realistic case of a system of many interacting particles,
particularly electrons. In order to obtain a tractable model, we
shall assume that the electrons interact via the electrostatic
mean field, their dynamics being described by a set of
one-dimensional (1D) hydrodynamic equations. As many experimental
studies on quantum information involve the manipulation of charged
particles, our approach may shed some light on the robustness of
such systems against perturbations from the environment.

{\it Model}.---The physical properties of our model are best
illustrated by considering its classical counterpart, the
so-called `cold plasma' model \cite{cold-plasma, bertrand}. In the
latter, the electron population is described by a phase-space
distribution function of the type: $f(x,v,t) =
n(x,t)~\delta[v-u(x,t)]$, where $n$ and $u$ are respectively the
electron density and average velocity, and $\delta$ denotes the
Dirac delta function. The support of such a distribution function
in the 2D phase space $(x,v)$ is a 1D curve defined by the
relation $v=u(x,t)$. The electron distribution evolves according
to the (collisionless) Vlasov equation. The ions are motionless
with uniform equilibrium density $n_0$ and periodic boundary
conditions (with $-L/2 \le x \le L/2$) are assumed for all
variables.

As long as there is no particle overtaking in the phase space,
each position $x$ corresponds to a well-defined velocity $u(x,t)$.
In this case, the Vlasov equation can be reduced to a closed set
of pressureless hydrodynamic equations:
\begin{eqnarray}
\label{eq:contin} \frac{\partial\,n}{\partial\,t} &+&
\frac{\partial\,(nu)}{\partial\,x} = 0, \\
\label{eq:force} \frac{\partial\,u}{\partial\,t} &+&
u\frac{\partial\,u}{\partial\,x} =
\frac{e}{m}\frac{\partial\,\phi}{\partial\,x},
\end{eqnarray}
where $-e$ an $m$ are respectively the electron charge and mass,
and $\phi(x,t)$ is the electric potential obeying Poisson's
equation
\begin{equation}
\label{eq:poisson} \frac{\partial^{2}\phi}{\partial\,x^2} =
\frac{e}{\varepsilon_0}\left(n - n_{0}\right).
\end{equation}
When particles overtake each other, the function $x \to u(x)$
becomes multivalued. The above hydrodynamic description then
breaks down, although the microscopic Vlasov model remains valid.

A trivial stationary solution of Eqs.
(\ref{eq:contin})-(\ref{eq:force}) is given by $n=n_0$, $u=0$. The
electron dynamics can be excited by modulating the initial
velocity: $u(x,t=0^+) = V_0 \cos(k_0 x)$, with $k_0=2\pi/L$, which
is equivalent to applying an instantaneous electric field at time
$t=0$. The classical dynamics is determined by a single
dimensionless parameter, namely the normalized wave number of the
initial perturbation $K_0 = k_0 V_0/\omega_p$, where $\omega_p =
(e^2 n_0/m \varepsilon_0)^{1/2}$ is the electron plasma frequency.
Note that $K_0$ can be viewed as the ratio of kinetic to potential
(electric) energy. It can be shown that two different regimes
exist. When $K_0<1$, electric repulsion dominates, so that the
electrons never overtake each other in the phase space. In this
case, the hydrodynamic equations
(\ref{eq:contin})-(\ref{eq:force}) can be solved analytically and
the solution displays nonlinear oscillations at the plasma
frequency. When $K_0>1$, the analytical solution breaks down and
the dynamics must be described by the microscopic Vlasov equation.
In the extreme case $K_0 \gg 1$, the dynamics becomes again
integrable, because it reduces to that of free-streaming
electrons. For moderate values of $K_0$ (but still larger than
unity), the electrons can be alternately free streaming and
trapped by the self-consistent (SC) potential. This regime
corresponds to the formation of complex vortices in the phase
space and leads to a chaotic dynamics, as was pointed out for the
similar scenario of nonlinear Landau damping \cite{valen}. In this
work, we will be mainly interested in the chaotic regime and use
the value $K_0=2$.

Quantum corrections to the hydrodynamic equations
(\ref{eq:contin})-(\ref{eq:force}) were previously derived
\cite{qfluid}. For fermions at zero temperature (a case that is
relevant to electrons in metals), the momentum conservation
equation (\ref{eq:force}) should be modified as follows:
\begin{equation}
\label{eq:force_q} \frac{\partial\,u}{\partial\,t} +
u\frac{\partial\,u}{\partial\,x} =
\frac{e}{m}\frac{\partial\,\phi}{\partial\,x}
+\frac{\hbar^2}{2m^2} \frac{\partial}{\partial
x}\left(\frac{\partial_x^{2}\sqrt{n}} {{\sqrt{n}}}\right) -
\frac{1}{mn} \frac{\partial\,P_F}{\partial\,x}.
\end{equation}
The second term on the right-hand side is the Bohm potential: this
is a dispersive term that prevents the breakdown of the quantum
hydrodynamics even when $K_0>1$ \cite{bertrand}. The third term is
the Fermi pressure, which in 1D can be written as: $P_F/P_0 =
(n/n_0)^3$, where the equilibrium pressure is given by the usual
formula, $P_0=\frac{2}{5}n_0 E_F$. $E_F$ is the Fermi energy
computed with the equilibrium density.

The continuity equation (\ref{eq:contin}) and the quantum momentum
equation (\ref{eq:force_q}) can be written in the form of a single
nonlinear Schr{\"o}dinger equation by introducing the effective wave
function $\Psi(x,t) = \sqrt{n(x,t)}\exp{(iS(x,t)/\hbar)}$, where
$S(x,t)$ is defined according to the relation $m u =
\partial_x S$, and $n=|\Psi|^2$. The wave function $\Psi$ obeys
the equation
\begin{equation}
\label{eq:nlse} i\hbar\frac{\partial\Psi}{\partial\,t} = -
\,\frac{\hbar^2}{2m}~\frac{\partial^2 \Psi}{\partial x^2} -
e\phi\Psi + \frac{3}{5} E_F \frac{|\Psi|^4}{n_0^2}~ \Psi.
\end{equation}

Equation (\ref{eq:nlse}) with Poisson's equation
(\ref{eq:poisson}) constitute the mathematical model used
throughout this Letter. The equilibrium Hamiltonian $H_0$ is time
dependent, as it depends self-consistently on the wave function,
but conserves both the total mass and the total energy. The
initial condition is analog to the classical one described in the
preceding paragraphs and can be easily derived from the velocity
perturbation $u(x,t=0^+)$ by using the relation between $S$ and
$u$. Two more dimensionless parameters (in addition to $K_0$)
intervene in the quantum model: (i) the normalized Planck constant
$h=\hbar \omega_p/m V_0^2$, which measures the importance of
quantum effects; and (ii) the normalized Fermi velocity $v_F/V_0$.
The latter affects very little the results (provided it is not too
large), and will be fixed at $v_F/V_0 = 0.1$ in the forthcoming
simulations.

\begin{figure}[htb]
\includegraphics[height=4.5cm]{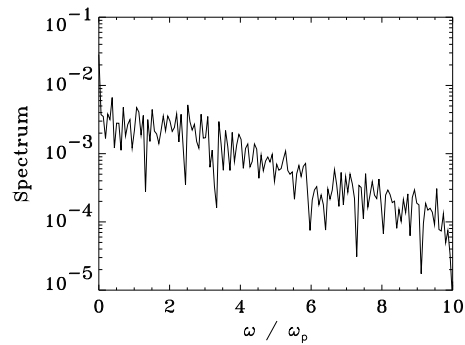}
\caption{\label{fig:fig1} Frequency spectrum of the potential
energy, for an unperturbed evolution with $K_0=2$ and $h=0.05$.}
\end{figure}

The numerical solution of Eq. (\ref{eq:nlse}) is obtained through
a splitting scheme that separates the kinetic and potential parts
of the Hamiltonian. Derivatives are computed with centered
differences. The resulting algorithm is second-order accurate both
in space and time.

{\it Results}.---First, we characterize the spectral properties of
the unperturbed Hamiltonian. We consider a case with $K_0=2$ and
$h=0.05$ and plot in Fig. 1 the frequency spectrum of the time
history of the electrostatic energy. The spectrum is broad and
virtually flat in the range $0 < \omega \lesssim 3 \omega_p$. The
dynamics is therefore sufficiently irregular to enable us to
compare our results to those obtained for a single-particle
chaotic Hamiltonian.

In order to study the behavior of the quantum fidelity, we need to
compare the evolution of $\Psi$ obtained with the unperturbed and
perturbed Hamiltonians. We use a static perturbation consisting of
a sum of a large number of uncorrelated waves: $\delta H(x)/mV_0^2
= \epsilon \sum_{j=N_{\rm min}}^{N_{\rm max}} \cos(k_j x
+\alpha_j)$, where $\epsilon$ is the amplitude of the
perturbation, $k_j = j~ (2\pi/L)$, and the $\alpha_j$ are random
phases. The wave number spectrum of the perturbation (i.e. the
values of $N_{\rm min}$ and $N_{\rm max}$) affect only very weakly
the behavior of the fidelity: therefore, we will focus our
analysis on the dependence of the fidelity on the amplitude
$\epsilon$.

\begin{figure}[htb]
\includegraphics[height=4.5cm]{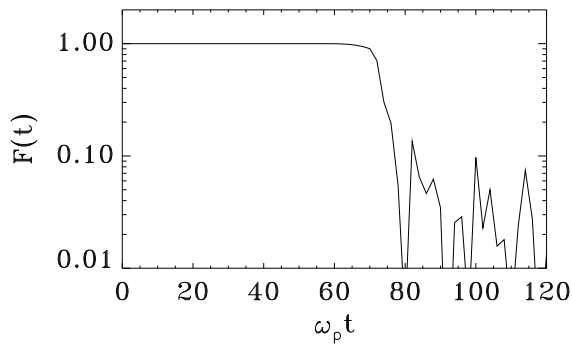}
\caption{\label{fig:fig2} Fidelity decay for $K_0=2$, $h=0.05$,
and perturbation $\epsilon =10^{-9}$.}
\end{figure}

A typical result for the quantum fidelity is presented in Fig. 2.
Contrarily to most single-particle cases, the fidelity does not
decay exponentially. Instead, it remains equal to unity until a
critical time $\tau_c$, after which it decays abruptly within a
few units of $\omega_p^{-1}$. This behavior is generic and was
observed for all set of parameters that were studied, provided the
dynamics is sufficiently irregular. Numerical tests showed that
the value of the critical time is independent on the time step.
However, as the underlying dynamics is chaotic, evolutions
computed using different time steps will inevitably diverge for
long times. Therefore, the details of the evolution for $t \gg
\tau_c$ are not quantitatively meaningful, and simply indicate
that the fidelity has dropped to very small values.

The observed drop in the quantum fidelity is related to a sudden
symmetry breaking of the wave function. Indeed, the evolution
equations (\ref{eq:poisson})-(\ref{eq:nlse}) for the unperturbed
Hamiltonian are invariant under the transformation $x \to -x$. If
the initial condition is an even function of $x$, this symmetry is
thus preserved in time, i.e. $\Psi(x,t) = \Psi(-x,t),~\forall t$.
But the perturbation $\delta H$ possesses no particular symmetry,
and one would expect that the symmetry of the initial state
quickly deteriorates. The symmetry properties can be conveniently
measured by the following quantity:
\begin{equation}
\Sigma(t) = ~\frac{2}{n_0 L} ~{\vline \int_0^{L/2} \Psi(x,t)
~\Psi^\star(-x,t) ~dx \vline}^{~2}, \label{eq:symm}
\end{equation}
which is equal to unity when $\Psi(x,t) = \Psi(-x,t)$. The
evolution of $\Sigma(t)$ for the perturbed Hamiltonian is plotted
in Fig. 3. The drop of the quantum fidelity happens virtually at
the same time as the breaking of the wave function symmetry. This
behavior is also generic across a wide range of parameters.

\begin{figure}[htb]
\includegraphics[height=4.5cm]{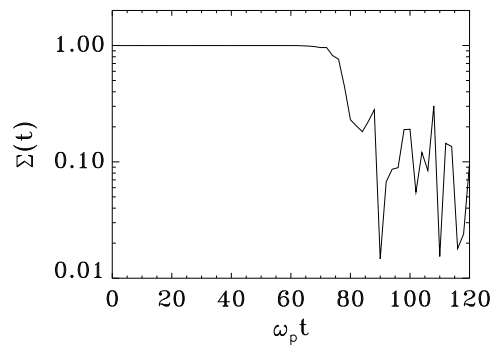}
\caption{\label{fig:fig3} Evolution of the symmetry $\Sigma(t)$
for the same case as in Fig. 2.}
\end{figure}

We further investigated the dependence of the critical time
$\tau_c$ on the perturbation amplitude $\epsilon$, for various
values of the normalized Planck constant $h$. The critical time is
defined as the time at which the fidelity has dropped to 10\% of
its maximum value, i.e. $F(\tau_c)=0.1$. Figure 4 shows that, for
small values of $h$, $\tau_c$ depends logarithmically on the
perturbation amplitude, i.e. $\tau_c \sim -t_0 \ln \epsilon$, with
$\omega_p t_0 \simeq 4.3$ (this is the straight line depicted in
Fig. 4). This logarithmic dependence appears to be universal (both
in slope and absolute value), at least for the values of $K_0$ and
$v_F/V_0$ adopted in these runs. For larger values of Planck's
constant ($h \gtrsim 0.2$), this behavior is less neat,
particularly for small perturbations.

\begin{figure}[htb]
\includegraphics[height=4.25cm]{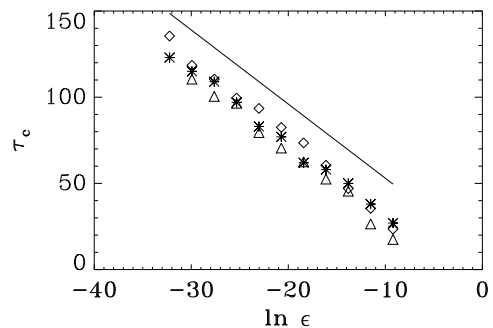}
\caption{\label{fig:fig4} Critical time $\tau_c$ (in units of
$\omega_p^{-1}$) versus perturbation amplitude $\epsilon$, for
$h=0.05$ (stars), $h=0.025$ (diamonds), and $h=0.0125$
(triangles). The solid line represents the curve $\tau_c \sim -t_0
\ln \epsilon$, with $\omega_p t_0 = 4.3$.}
\end{figure}

A similar pattern was observed for a chaotic quantum map
\cite{casati}: in that case, the fidelity stays equal to unity
until a critical time, after which it starts to decay
exponentially at the classical Lyapunov rate. No sudden drop was
observed, as is the case for our simulations.

In order to better evaluate the impact of the SC field, we
performed some simulations where all nonlinear terms have been
suppressed in Eqs. (\ref{eq:poisson})-(\ref{eq:nlse}). The first
nonlinearity comes from the Fermi pressure and can be removed
simply by setting $E_F=0$. The SC nonlinearity comes from the fact
that the electric potential depends on the wave function through
the electron density $n=|\Psi|^2$. To remove this nonlinearity, we
define the electron density independently of $\Psi$, as the sum of
traveling plane waves: $n=n_{\rm ext} \equiv n_0[1+\delta
\sum_{j=1}^{25} k_j^2 \cos(k_j x -\omega_p t+\alpha'_j)]$, where
$\delta$ is the amplitude of the density fluctuations, and the
$\alpha'_j$ are random phases. This definition is plugged into
Poisson's equation to yield the electric potential. As the
resonances of the plane waves overlap in phase space, the
resulting (time-dependent) Hamiltonian $H_0$ is likely to display
chaotic regions \cite{reson}. The fidelity decay is studied by
perturbing the Hamiltonian in the same way as in the SC case.

In Fig. 5 we plot the quantum fidelity for $\delta=0.5$,
$h=0.025$, and several values of the perturbation $\epsilon$. The
fidelity decay is exponential and begins at $t=0$. The decay rate
is approximately proportional to the square of the perturbation,
which shows that we are in the so-called Fermi golden rule regime
\cite{jacquod,casati}. However, contrarily to Ref. \cite{casati},
no plateau was observed for short times.

\begin{figure}[htb]
\includegraphics[height=4cm,width=6cm]{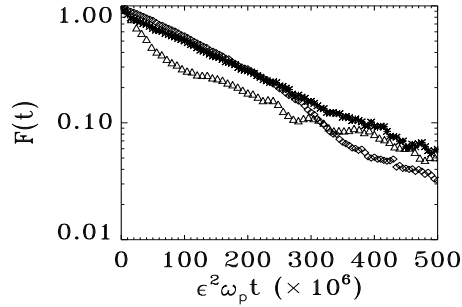}
\caption{\label{fig:fig5} Fidelity decay for a purely external
Hamiltonian: $\epsilon=10^{-3}$ (stars); $\epsilon=2 \times
10^{-3}$ (triangles); $\epsilon=5 \times10^{-4}$ (diamonds). Time
is rescaled to the square of the perturbation $\epsilon$.}
\end{figure}

\begin{figure}[htb]
\includegraphics[height=4cm,width=6cm]{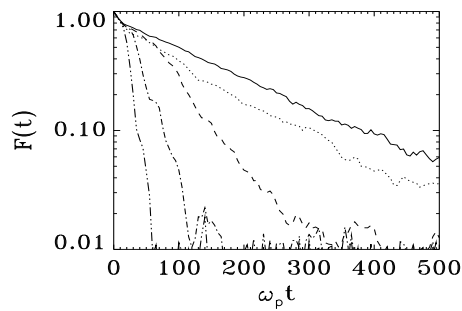}
\caption{\label{fig:fig6} Fidelity decay for a mixed external and
SC Hamiltonian, for $\beta=0$ (solid line); $\beta=0.01$ (dotted);
$\beta=0.03$ (dashed); $\beta=0.1$ (dot-dashed); $\beta=0.3$
(dot-dot-dashed).}
\end{figure}

Finally, we studied a case where both the SC and the external
fields are present. This is accomplished by defining the electron
density as: $n(x,t)= n_{\rm ext} + \beta (|\Psi|^2-n_0)$. By
varying $\beta$ and $\delta$, we can move continuously from a
purely `external' regime ($\beta=0$) to a purely SC one
($\delta=0$, $\beta=1$). We concentrate on a case with
$\delta=0.5$, $h=0.025$, and $\epsilon=10^{-3}$, and vary the
value of $\beta$ (Fig. 6). For short times, the fidelity decays
exponentially with the same rate as in the purely `external' case.
Subsequently, the decay becomes faster, and for large values of
$\beta$ we almost recover the abrupt drop of Fig. 2.

{\it Discussion}.--- The present work is a first attempt at
studying quantum fidelity decay in a system of electrons
interacting through their SC electric field. Our numerical results
show that the quantum fidelity can display a rapid decrease. Such
effect is probably related to the fact that the unperturbed
Hamiltonian $H_0$ depends on the wave function. When the
perturbation $\delta H$ induces a small change in $\Psi$, $H_0$ is
itself modified, which in turns affects $\Psi$, and so on. Because
of such nonlinear loop, the perturbed and unperturbed solutions
can diverge very fast (typically, within a few $\omega_p^{-1}$).
In contrast, for the single-particle dynamics, $H_0$ is fixed and
the solutions only diverge because of the small perturbation
$\delta H$.

In summary, it appears that the `natural' response of a SC system
to a perturbation is a sudden drop of the fidelity after a
quiescent period rather than an exponential decay. Further studies
will be needed to understand whether these results extend to more
complex models, going beyond the mean-field approach adopted here.

We acknowledge fruitful discussions with R. A. Jalabert.
%

\end{document}